\documentclass[aps,twocolumn,epsf,eps,floats,pre,nofootinbib,showpacs]{revtex4-1}
\usepackage[usenames]{color}
\usepackage{amssymb,amsfonts,amsmath}
\usepackage{acronym}
\usepackage{graphicx}
\usepackage{ifpdf}
\ifpdf
\DeclareGraphicsExtensions{.pdf,.eps,{}}
\else
\fi

\begin{document}

\title{Static correlations functions and domain walls in glass-forming
  liquids: the case of a sandwich geometry}

\author{Giacomo Gradenigo} 
\email{ggradenigo@gmail.com}
\affiliation{Dipartimento di Fisica, Universit\`a\ Sapienza, P.le
  Aldo Moro 2, 00185 Roma, Italy}
\affiliation{Istituto Sistemi Complessi (ISC), Consiglio Nazionale delle Ricerche (CNR), UOS Sapienza, Via
  dei Taurini 19, 00185 Roma, Italy} 

\author{Roberto Trozzo}
\affiliation{Dipartimento di Fisica, Universit\`a Sapienza - P.le
  A. Moro 2, 00185, Roma, Italy}

\author{Andrea Cavagna}
\affiliation{Istituto Sistemi Complessi (ISC), Consiglio Nazionale
  delle Richerche (CNR), UOS Sapienza, Via dei Taurini, 19, 00185
  Roma, Italy}
\affiliation{Dipartimento di fisica, Universit\'a Sapienza, P.l Aldo
  Moro, 2, 00185 Roma, Italy}

\author{Tom\'as S.\ Grigera}
\affiliation{Instituto de Investigaciones
  Fisicoqu{\'\i}micas Te{\'o}ricas y Aplicadas (INIFTA) and
  Departamento de F{\'\i}sica, Facultad de Ciencias Exactas,
  Universidad Nacional de La Plata, c.c. 16, suc. 4, 1900 La Plata,
  Argentina}
\affiliation{CONICET La Plata, Consejo Nacional de
  Investigaciones Cient{\'\i}ficas y T{\'e}cnicas, Argentina}

\author{Paolo Verrocchio} \affiliation{Dipartimento di Fisica and
  Interdisciplinary Laboratory for Computational Physics (LISC),
  Universit{\`a} di Trento, via Sommarive 14, 38050 Povo, Trento,
  Italy} \affiliation{Istituto Sistemi Complessi (ISC-CNR), UOS
  Sapienza, Via dei Taurini 19, 00185 Roma, Italy}
\affiliation{Instituto de Biocomputaci\'on y F\'{\i}sica de Sistemas
  Complejos (BIFI), Spain}

\begin{abstract}
  The problem of measuring nontrivial static correlations in deeply
  supercooled liquids made recently some progress thanks to the
  introduction of amorphous boundary conditions, in which a set of
  free particles is subject to the effect of a different set of
  particles frozen into their (low temperature) equilibrium positions.
  In this way, one can study the crossover from nonergodic to ergodic
  phase, as the size of the free region grows and the effect of the
  confinement fades. Such crossover defines the so-called point-to-set
  correlation length, which has been measured in a spherical geometry,
  or cavity.  Here, we make further progress in the study of
  correlations under amorphous boundary conditions by analyzing the
  equilibrium properties of a glass-forming liquid, confined in a
  planar (``sandwich'') geometry.  The mobile particles are subject to
  amorphous boundary conditions with the particles in the surrounding
  walls frozen into their low temperature equilibrium configurations.
  Compared to the cavity, the sandwich geometry has three main
  advantages: i) the width of the sandwich is decoupled from its
  longitudinal size, making the thermodynamic limit possible; ii) for
  very large width, the behaviour off a single wall can be studied;
  iii) we can use ``anti-parallel'' boundary conditions to force a
  domain wall and measure its excess energy.  Our results confirm that
  amorphous boundary conditions are indeed a very useful new tool in
  the study of static properties of glass-forming liquids, but also
  raise some warning about the fact that not all correlation functions
  that can be calculated in this framework give the same qualitative
  results.
\end{abstract}

\maketitle

\acrodef{RFOT}{Random First Order Transition}
\acrodef{ABC}{Amorphous Boundary Conditions}
\acrodef{PBC}{Periodic  Boundary Conditions}
\acrodef{ARS}{Aspect-Ratio Scaling}
\acrodef{RFIM}{Random Field Ising Model}

\section{Introduction}
\label{sec:intro}

The sharp slowdown observed in supercooled liquids at low temperatures
has long been conceptually connected to the buildup of structural
(static) correlations.  Yet, due to the amorphous nature of the
excitations, it has proved rather difficult to identify them and to
measure their size.  For this reason, dynamical correlations
\cite{heterogeneities:sillescu99, review:Ediger00, ISRAELOFF00,
  berthier03b, Cugliandolo03, BERTHIER03, BERTHIER04,
  heterogeneities:sillescu02, heterogeneities:garrahan02,
  heterogeneities:Berthier05, heterogeneities:berthier07a,
  heterogeneities:berthier07b} were detected much before static ones,
and only recently were structural correlations unveiled, using novel
techniques \cite{review:kivelson97,self:prl07,
  landscape:widmer-cooper08, glassthermo:tanaka10,
  glassthermo:coslovich2011}.

Among these techniques, numerical simulations with \acfp{ABC}, and the
related point-to-set correlation length $\xi$, have proved very
fruitful \cite{Scheidler02, self:prl07, self:nphys08,
  confinement:berthier12, confinement:kob12,
  correlation-length:hocky12, dynamics:montanari06}.  Implementing
\acp{ABC} is simple, at least in numeric simulations.  Consider a set
of {\em mobile} and another one of {\em frozen} particles and let the
mobile particles evolve under the influence of the frozen ones,
eventually reaching thermodynamic equilibrium.  The simplest case is
when the frozen particles belong to a single equilibrium configuration
surrounding a spherical cavity of mobile particles, of radius $R$.  It
is possible then to define an overlap $q(R)$ and to measure the
similarity at the centre of the sphere between two configurations.
The dependence of the overlap on the linear size $R$ of the cavity
yields the correlation lenght $\xi$, defined by the crossover at $R
\sim \xi$ among the values $q(R) \sim 1$ (almost identical
configuration) and $q(R) \sim 0$ (statistically independent
configurations).

\begin{figure}
  \includegraphics[width=\columnwidth]{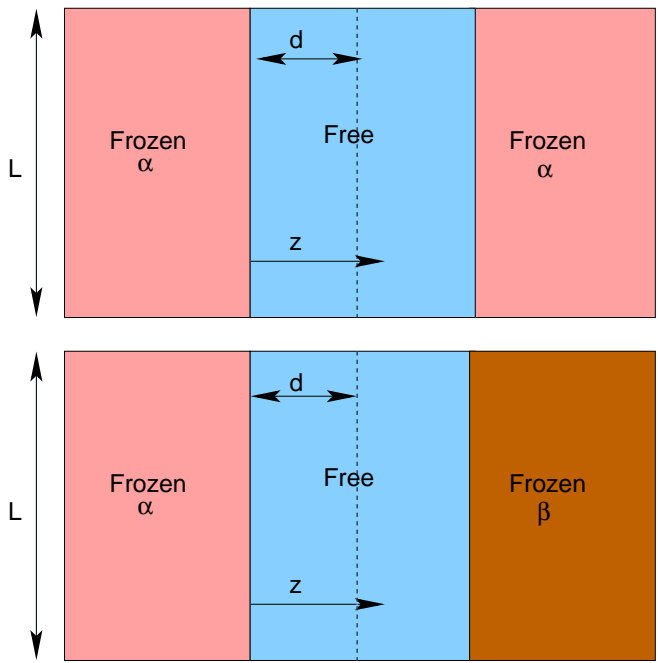}
  \caption{Cartoon of the sandwich geometry.  In the ``paralell'' (or
    $\alpha\alpha$) setup (top) both frozen walls are taken
    from the same equilibrium configuration, while in the
    ``anti-parallel'' ($\alpha\beta$) case, they come from different
    configurations (bottom).}
  \label{fig:cartoon}
\end{figure}

The original \ac{ABC} spherical realization can be generalized to
different geometries, where the frozen particles do not necessarily
form a closed cavity \cite{confinement:kob12, confinement:berthier12,
  mosaic:zarinelli10}.  In this work we study the case of a planar (or
``sandwich'') geometry (see Fig.~\ref{fig:cartoon}).  As with the
spherical geometry, in the sandwich we can calculate a point-to-set
length by studying the sandwich width beyond which the internal mobile
particles reach ergodicity.  In this respect our study aims to verify
the results obtained in the cavity and test their robustness.  In
particular, we are interested to check whether or not the anomalous
nonexponential behaviour of the point-to-set correlation function at
low $T$ observed in the spherical geometry \cite{self:nphys08} is also
found in the sandwich.  To check how general is this nonexponential
behaviour is important becoause it is one of the very few qualitative
thermodynamic landmarks of the deeply supercooled phase.

But the sandwich also allowd us to study cases that are out of reach
in the cavity.  First, in the sandwich the width $d$ and the
longitudinal size $L$ are independent parameters, so that we can
perform the limit limit $L\to\infty$ while the confinement length
keeping $d$ finite.  This thermodynamic limit is clearly impossible in
the cavity.  This limit is important, as by increasing the number of
mobile particles at constant degree of confinement, we can check
whether or not the finite-size crossovers of the correlation functions
turn into {\it bona fide} transitions.

Second, when the two walls are very far from each other
($d\to\infty$), we can study the decay of the overlap off a single
wall, as a function of the distance $z$ from the wall.  This is not
strictly impossible in spherical geometry, but in that case one could
be exposed to spurious curvature effects that are absent in the
sandwich.

Third, in the planar geometry we can use {\it different} amorphous
boundary conditions on the two sides of the sandwich
(Fig.~\ref{fig:cartoon}, bottom), which is also impossible in the
cavity.  This sort of ``anti-parallel'' boundary conditions can be used
to force a domain wall in the system, and therefore to measure its
excess energy and the stiffness exponent $\theta$.  These quantities
are crucial in any phenomenological description of the glass
transition, so that any new tool able to provide information on these
quantities may be helpful.

\section{Model and simulation details}
\label{sec:model}

We study the soft-sphere binary mixture \cite{soft-spheres:bernu87}, a
simple model of supercooled liquids widely studied before, and in
which the point-to-set correlation has been computed using a cavity.
We use the accelerated Swap Monte Carlo algorithm
\cite{algorithm:Grigera01} to thermalize the system at temperatures as
low as possible.  We run simulations at $T=0.482, 0.350, 0.246,0.202$.
The first two temperatures correspond to the high-temperature liquid,
the third is near the ``onset'' or ``landscape-influenced''
temperature \cite{landscape:brumer04} and the lowest temperature lies
in the supercooled regime, in which the landscape is dominated by
minima of the potential energy rather than saddle points.

The confined system is generated from configurations taken from
equilibrated periodic-boundary-conditions runs.  These runs were done
with density $\rho=1$ and box sizes $L=16$ and $L=25.3$.  At each
temperature we then chose several (from 16 to 24) configurations and
artificially froze in their equilibrium positions all but $M$
particles contained within a region of the simulation box in the shape
of a box of size $2d \times L^2$ (we measure $d$ along the $z$ axis).

In order to keep the density fixed within the region of mobile
particles, it is a standard practice to place \emph{virtual} walls at
the border of such mobile regions.  What we do is the following:
taking configurations of the liquid system, we place a hard wall
potential enclosing the free particles.  This destroys translational
invariance along the $z$ axis, but not along the $xy$ planes, creating
a \emph{sandwich} of mobile particles surrounded by two infinite walls
of frozen liquid.

The main observable we consider is the infinite time limit of the
local density-density correlations.  More precisely, we define the
\emph{overlap}  $q(z;d)$, as follows: we partition the simulation
box in many small cubic boxes of side $\ell$, such that the
probability of finding more than one particle in a single box is
negligible.  If $n_i$ is the number of particles in box $i$, then,
\begin{equation}
  q(z;d) = \lim_{t\to\infty} \frac{1}{\ell^3 N_i} 
  \sum_{i\in v} \langle n_i(t_0)  n_i(t_0+t)  \rangle,
  \label{eq:overlap-def}
\end{equation}
where the sum runs over all boxes that lie on a plane parallel to the
$xy$ plane at the given distance $z$ from one reference wall, 
$N_i$ is the number of boxes in each of those planes, and 
$\langle\ldots\rangle$ indicates a thermal average. Normalization 
is such that the overlap of two identical configurations is $1$ on 
average, while for totally uncorrelated configurations $q_=
q_0 \equiv \ell^3=0.062876$.

\section{Different static lengthscales}
\label{sec:lengths}

In this section we study the the overlap, Eq.~\ref{eq:overlap-def}, in
the sandwich geometry described above, in which mobile particles are
confined within a volume $2d L^2$ by two walls made of frozen
particles (top scheme in Fig.~\ref{fig:cartoon}). In our description,
$d$ is the half-width of the sandwich; we believe this is the correct
variable to compare our results (especially lengthscales) with the
spherical cavity case.  It is important to note that both walls are
made from particles taken from the \emph{same} equilibrated
configuration.

In general, the overlap is a measure of the nonergodicity of the
mobile part of the sandwich due to the frozen boundary conditions.
When the overlap is nonzero (more precisely: larger than its ergodic
value $q_0$) it means that the phase space available to the particles'
relaxation is reduced by the confinement.  It is therefore natural to
ask how ``far away'' the walls need to be so that ergodicity is
restored.  In the case of the sandwich, this question can be asked in
two ways:
\begin{enumerate}
\item \emph{How big} must the wall separation be so that the
  liquid inside behaves like the bulk?
\item Given a very large (or infinite) cavity, \emph{how far} from the
  walls must one look so that the liquid behaves like in the bulk?
\end{enumerate}     
The first question implies that one is observing the overlap \emph{as
  a function of $d$} at some reference position within the sandwich
(typically at the center, since influence of the walls very near the
interface is always expected).  In the second question, one considers
the overlap \emph{as a function of $z$} at fixed, very large $d$. As
we shall see, the two questions have qualitatively and quantitatively
different answers.

\subsection{Point-to-set correlation length}

\begin{figure}
  \includegraphics[width=\columnwidth]{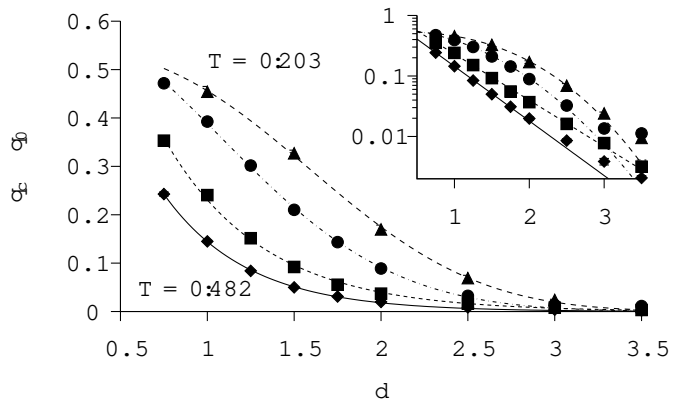}
  \caption{Overlap at the center of the sandwich vs.\ sandwich
    half-width $d$ in the parallel setup for (from left to right)
    $T=0.482$, 0.350, 0.246, 0.202.  Lines are exponential or near the
    centercompressed-exponential fits (see text).  Inset: same data in
    semilog plot.}
  \label{fig:qvsd-alpha-alpha}
\end{figure}

We first study the decay of the overlap following the point-to-set
prescription, i.e.\ measuring the overlap at the center of the
sandwich ($z=d$) and varying the distance $d$ between the walls (by
symmetry, we can actually average the overlap over the whole central
plane).  We call this point-to-set overlap, computed at the central
plane $q_c(d)$.  The behaviour of this quantity is shown in
Fig.~\ref{fig:qvsd-alpha-alpha} for four different temperatures.  The
scale of decay of this function defines the point-to-set correlation
length $\xi$.

A notable feature of $q_c(d)$ is that its decay crosses over from simple 
exponential at high temperatures to non-exponential at low temperatures.
In the low $T$ phase a simple exponential fit does a very bad job, while 
the curves can be fitted via a ``compressed exponential'' form,
\begin{equation}
q_c(d) = \Omega \exp [-(d/\xi)^{\zeta}]  +q_0 \ ,
\label{compressed}
\end{equation}
where the anomaly exponent $\zeta$ measures the deviation from
exponentiality.  This specific form is by no means the only one
capable of capturing the nonexponential shape.  The relevant point is
that such nonexponential behaviour is present, and that it is useful
to have a scalar parameter (in this case $\zeta$) to quantify it.

At high temperatures a semilog plot shows that the curves are
reasonably exponential, so in order to avoid overfitting we fix
$\zeta=1$ and we fit the data to a pure exponential.  On the other
hand, at low temperatures there is a clear deviation from
exponentiality (inset of Fig.~\ref{fig:qvsd-alpha-alpha}), so that the
nonexponential fit (Eq.~\ref{compressed}) is used.  At the lowest $T$
we obtain $\zeta=2.7 \pm 0.2$ (see Table 1 for all values of $\zeta$).
 
This progressive sharpening of the decay at low temperatures (growing
of the anomaly exponent) is also found in the spherical cavity
\cite{self:nphys08}, but the numerical value of the exponent $\zeta$
is different (lower) in the sandwich case.  Thus the geometry of the
system may influence the strength of the exponential/non-exponential
crossover but the \emph{existence} of the crossover itself seems not
to depend on the geometry and it is therefore a robust result.

\subsection{Penetration length}
\label{sec1:lengths}

\begin{figure}
  \includegraphics[width=\columnwidth]{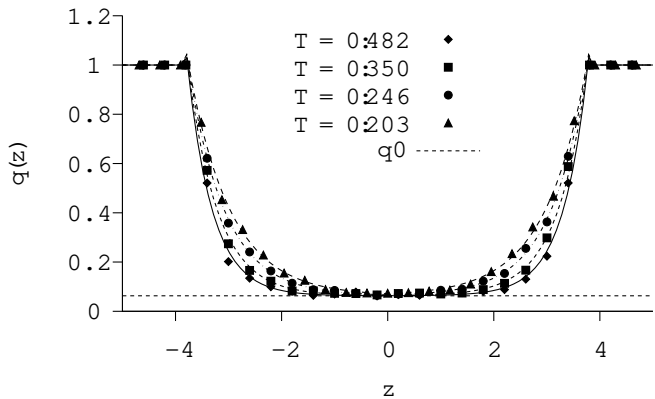}
  \caption{Overalp in the paralell setup as a function of $z$ for a
    sandwich of half-width $d=4$ at several temperatures.  The free
    region is wide enough that the overlap can reach its bulk value
    $q_0$ near the center.}
  \label{fig:qvsz-double}
\end{figure}

We now consider the decay of the overlap off one single wall.  It is
clear from Fig.~\ref{fig:qvsz-double} that for a large enough
values of the sandwich width $d$, the overlap has enough room to decay
to its liquid value $q_0$ at the central plane, at all
temperatures. Therefore, the decay of the overlap, $q(z,d\gg\xi)$, as
a function of the distance $z$ from one of the two walls, is perfectly
equivalent to the decay of the overlap from a single wall in a
semi-infinite geometry.  We call this quantity simply $q(z)$.

\begin{figure}
  \includegraphics[width=\columnwidth]{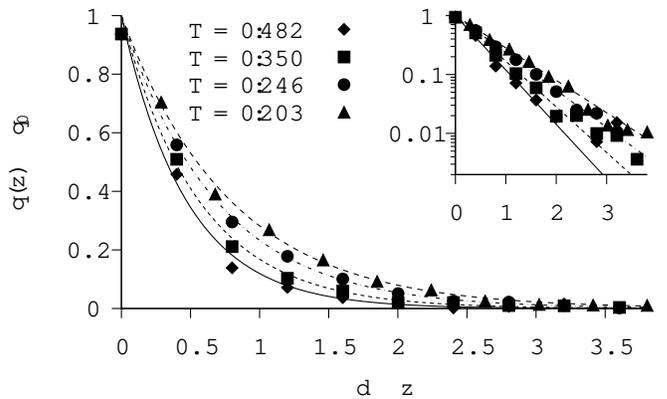}
  \caption{Overlap vs.\ $d-z$ (distance from wall) at fixed $d$ (same
    data as Fig.~\ref{fig:qvsz-double}) with pure exponential fits.
  Inset: same data in semilog scale.}
  \label{fig:qvsz}
\end{figure}

Fig.~\ref{fig:qvsz} shows the behaviour of $q(z)$ focusing on one
single wall.  The first feature that we notice, in comparison with the
point-to-set correlation function, is that at all temperatures data
are well fitted with a simple exponential,
\begin{equation}
q(z)=\exp[ - z/\lambda ]  +q_0 \ ,
\label{eq:exp-decay}
\end{equation}
where $\lambda$ is the penetration length.  Independently from the fit
quality, the pure exponential behaviour is evident from the semilog
plot (inset of Fig.~\ref{fig:qvsz}).  This result is in agreement with
the results obtained with a single wall in \cite{confinement:kob12,
  confinement:berthier12}: an exponential decay at all temperatures
with no sign of crossover to non-exponentiality.  This feature is a
remarkable difference with respect to the nonexponential point-to-set
correlation $q_c(d)$.  Such difference is perhaps not surprising: the
two quantities are conceptually not the same, as we shall argue in the
next section.

Apart from the functional form of the decay, another difference that
we immediately notice is that the penetration length $\lambda$ seems
to be smaller than the correlation length $\xi$.  We will return to
this in Sec.~\ref{sec:lengthsc-comp-ising}.

\section{Mosaic in the sandwich}
\label{RFOTsandwich}

\subsection{Naive argument}

We have shown in the previous subsection that the decay of the overlap
in the sandwich has the same exponential/super-exponential crossover
with temperature that is observed in the spherical geometry.  Such
anomalous non-exponential behaviour at low $T$ was explained in
~\cite{self:nphys08} by using a generalization of the RFOT framework.
In this Section we will show that, at least at the naive level, the
same RFOT arguments that hold in the cavity can be also applied to the
present sandwich geometry.

The basic idea of RFOT is that the relaxation of a confined system is
regulated by trade-off between a cost and a gain of exploring states
other from the one fixed into the amorphous boundary conditions. The
cost is the free energy the system has to pay to form an interface
when it changes state, whereas the gain is the entropic surplus the
system enjoys by changing state \cite{mosaic:kirkpatrick87,
  mosaic:kirkpatrick89, mosaic:bouchaud04}.  The slight complication
of the sandwich is that one must careful to take account of its
anisotropic geometry.  Unless we are at some very specific value of
the parameters (that we shall discuss later), it seems reasonable to
assume that the rearrangement of the mobile part of the sandwich
happens independently within uncorrelated regions, whose longitudinal
size is larger that the correlation length $\xi$. If we call $A$ and
$B$ two such regions, we are saying that
\begin{equation}
\mathcal{Z}_{A+B} \approx \mathcal{Z}_A\mathcal{Z}_B   .
\label{pongo}
\end{equation}
This means that the overlap of the mobile particles will be basically
a longitudinal average of the overlaps of such uncorrelated regions,
\begin{equation}
 q_c(d,L)  = \frac{1}{n}\sum_{r=1}^{n} q_c^{(r)}(d)  ,
\end{equation} 
where $n$ is the number of uncorrelated regions along the sandwich.

If we accept this, then the RFOT argument can be run over one
independent region of longitudinal size $\sim \xi$, and of width $\sim
d$.  Exactly as in the cavity, the entropic gain is (all relations are
given in the three dimensional case)
\begin{equation}
\Delta F_\mathrm{gain} \sim T\Sigma  \xi^2 d   .
\end{equation}
The surface tension cost, however, is trickier than in the cavity.  On
one hand, we know that it must scale like a length to
the power $\theta$, the stiffness exponent.  On the other hand, we also
expect from extensivity reasons that this cost must scale like the
longitudinal size of the rearranging region to the power $d-1=2$:
surely, if we build a super-sandwich by putting many sandwiches aside,
the total cost must be additive.  We can encapsulate these two
requirements by writing
\begin{equation}
\Delta F_\mathrm{cost} \sim Y d^\theta f(d/\xi)  ,
\end{equation}
where $f(d/\xi)$ is a scaling function that, due to extensivity, must obey the relation
\begin{equation}
f(d/\xi)\sim \xi^2/d^2 ,  \qquad \xi\gg1.
\end{equation}
In the end, we get
\begin{equation}
\Delta F_\mathrm{cost} \sim Y \xi^2 d^{\theta-2}   .
\end{equation}
As usual in the RFOT argument, we obtain the correlation length, i.e.\
the lengthscale at which the overlap decays to zero, as the
value of $d$ where the two contributions balance, $\Delta F_\mathrm{gain}\sim
\Delta F_\mathrm{cost}$. This yields
\begin{equation}
d_\mathrm{RFOT} \sim \left(\frac{Y}{T\Sigma}\right)^\frac{1}{3-\theta}   .
\end{equation}
This is the same prediction as RFOT gives in a cavity geometry. This
sharp RFOT scenario should then be smoothed by including the surface
tension fluctuations, following Ref.~\cite{self:nphys08}.  In this way
one gets a $q_c(d)$ that decays on a scale $d_\mathrm{RFOT}$, and
whose decay is sharper and sharper (larger exponent $\zeta$) the lower
the temperature, in agreement with what we find numerically. In this
context, the point-to-set correlation length $\xi$ must be identified
with the RFOT lengthscale $d_\mathrm{RFOT}$,
\begin{equation}
\xi\sim d_\mathrm{RFOT}   .
\end{equation}
Note that, at the level of this naive treatment, the difference
between the two walls vs.\ the single wall geometry, and therefore the
difference between $q_c(d)$ and $q(z)$, is quite clear.  In the single
wall case the entropic gain is infinite, as flipping the entire
semi-plane is an advantage over {\it any} interface energy.  So, we do
not expect any trade-off in that case.  However, even in the single
wall geometry, the best distance $\lambda$ where to locate the
interface will be nontrivial, since it may be entropically
inconvenient for the system to squeeze the interface too close to the
wall.  But it will be the entropy of the rough {\it interface}, not
that of the {\it bulk}, to matter.  For this reason, we do not expect
the growth of $\lambda$ to be regulated by a classic RFOT trade-off,
while do we expect so for the point-to-set length $\xi$.  No surprise,
then, that the two quantities are different.

\subsection{Sharpening in the thermodynamic limit?}

As we argued before, an advantage of the sandwich geometry over the
cavity is that one can tune the width $d$ and the longitudinal size
$L$ independently.  This means that (at least in principle) in the
sandwich one can perform the thermodynamic limit $L\to\infty$ at {\it
  fixed} $d$. However, because of the statistical factorization
hypothesis (Eq.~\ref{pongo}), the longitudinal size $L$ plays no role
at all in the RFOT argument.  In general, this is not necessarily
correct.  It has been argued in Ref.~\cite{pinning:cammarota12} that,
depending on the specific system's geometry and on the dimensionality,
the limit $L\to\infty$ can actually turn the $q_c(d)$ smooth decay
with $d$, into a {\it bona fide}, sharp transition at $d =
d_\mathrm{PTS}$, even at $T>T_k$.  We will only sketch the argument
here.

What we have disregarded above is the interaction between the
different rearranging regions in the mobile part.  Consider two
neighbouring regions, $A$ and $B$, and ask which is the propensity of
$A$ to decorrelate from its initial state.  Clearly, this depends on
the frozen boundaries enclosing $A$, but also on the state of the
neighbouring particles in $B$~\cite{glass:kurchan11,
  mosaic:cammarota12b}.  The state of $B$ may favour or not the
ergodization of $A$~\cite{glass:kurchan11, mosaic:cammarota12b}, and
one should take into account this interaction.  According to the
theoretical scenario of Ref.~\cite{pinning:cammarota12}, it turns out
that {\it exactly at the transition point} $d=d_\mathrm{RFOT}$, this
longitudinal interaction can make the sandwich long-range correlated
along the longitudinal plane.  This phenomenon would work in the
direction of making the transition between IN and OUT states sharper
and sharper.  However, in \cite{pinning:cammarota12} it is also
remarked that such transition is smoothed by the presence of the
disorder (disorder is generated by the surface tension fluctuations
along the sandwich) and that this has the effect to suppress the
transition in a $d=3$ sandwich, which is our case.  Therefore, one
should not expect any particular effect when increasing $L$ (the
transition would not be suppressed in a $d=4$ sandwich, nor in a $d=3$
system with randomly frozen particles, though --- see
Ref.~\cite{mosaic:cammarota12b}).

\begin{figure}
  \includegraphics[width=\columnwidth]{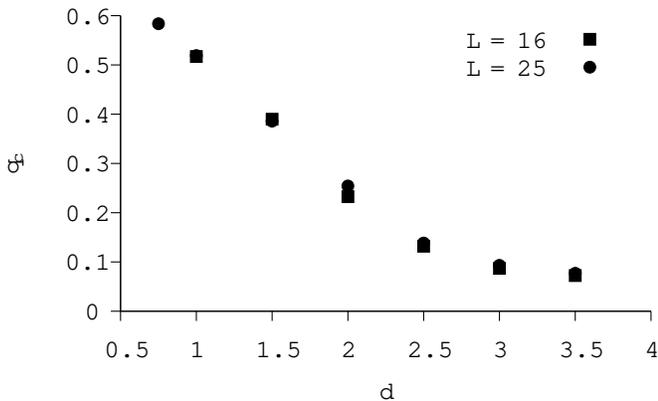}
  \caption{Overlap at center vs.\ sandwich half-width $d$ at $T=0.203$
  and two values of $L$.}
  \label{fig:qfss}
\end{figure}

We report the overlap $q_c(d,L)$ for two different sizes, $L=16$ and
$L=25$, in Fig.~\ref{fig:qfss}.  Indeed we do not find any evidence of
a sharpening of the decay of for larger $L$, which confirms the
expectation above.  For a $3$-dimensional sandwich, thus, the naive
RFOT argument provided at the beginning of this Section is probably
good enough.

\section{``Anti-parallel'' boundary conditions}
\label{sec:energy}

We now turn to the study of the excess energy produced by forcing an
interface in the mobile part of the sandwich.  To do this we use
``anti-parallel'' boundary conditions: we freeze particles on one wall
in a configuration $\alpha$, and those on the other wall in a
different configuration $\beta$ (see Fig.~\ref{fig:cartoon}, bottom).
The reason to study this geometry is twofold.  First, the surface free
energy cost is a key ingredient of the RFOT theory
\cite{kirkpatrick89, mosaic:bouchaud04}, but little is known about it.
The very possibility of measuring the surface tension between
amorphous states is at present under debate \cite{mosaic:franz11}, and
recently arguments against the existence of amorphous domain walls (a
question which is closely related to the previous one) across the bulk
of the glassy liquid have been proposed~\cite{mosaic:cammarota12b}.
Moreover it is not clear if the surface tension in a supercooled
liquid is a purely entropic phenomenon or if it also includes an
energy part due to the mismatch of different states.

Secondly, the numerical study of the excess energy provides in
principle a method to estimate the stiffness exponent $\theta$,
another crucial player in the RFOT formulas, regulating the growth of
the correlation length.  Unfortunately, we shall see that, although
the sandwich geometry is in principle ideal to determine $\theta$
through the technique of the aspect ratio scaling, in practice the
present values of the correlation length are not large enough to make
an unambiguous determination of $\theta$.

\subsection{Interface energy}

\begin{figure}
  \includegraphics[width=\columnwidth]{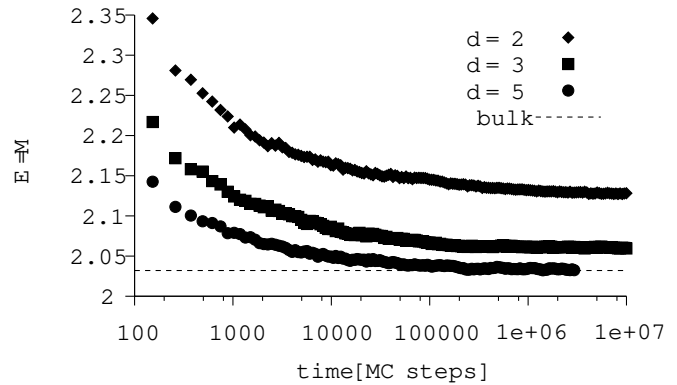}
  \caption{Excess energy per mobile particle at $T=0.246$ and several
    values of $d$, together with the bulk (PBC) average value.}
  \label{fig:Evst}
\end{figure}

We define the excess energy as the difference between the {\it
  extensive} energy of the mobile part of the sandwich with
``anti-parallel'' boundary conditions and ``parallel'' boundary
conditions,
\begin{equation}
\Delta E(d) = E_{\alpha\beta}(d) - E_{\alpha\alpha}   ,
\end{equation}
averaged over $16$ samples.  It is important to note that, if we
average over a large enough number of samples, the energy
$E_{\alpha\alpha}$ is equal to the (extensive) equilibrium energy.  In
the anti-parallel case, relaxation of the energy is in general {\it
  very} slow, and it get slower for smaller $d$.  This fact is true
even using the accelerated swap dynamics (Fig.~\ref{fig:Evst}).
Clearly, the system is unhappy with the $\alpha\beta$ boundary, likely
because of the forcing of a domain wall.  For this reason, at the
smallest values of $d$ we do not reach a plateau of the energy even
for our longest time.  In these cases we extrapolate the limiting
value of the excess energy by using a power-law fit,
\begin{equation}
\Delta E(t,d) = \Delta E(d) + At^{-\alpha}  .
\label{extrapolation}
\end{equation}

\begin{figure}
  \includegraphics[width=\columnwidth]{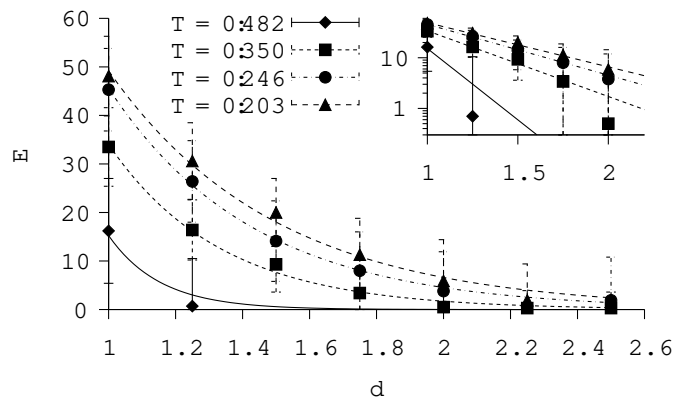}
  \caption{Excess energy vs. $d$ at several temperatures and $L=16$.
    Lines are exponential fits.  Inset: same data in semilog scale.}
  \label{fig:DEvsT}
\end{figure}

Fig.~\ref{fig:DEvsT} shows the excess energy $\Delta E(d)$ for all the
values of $d$ studied.  As expected, $\Delta E$ decays when increasing
$d$, and, at fixed $d$, the excess energy grows upon lowering the
temperature. At the largest temperature $T=0.482$ though, $\Delta E$
is basically always zero except for the smallest $d$.  The fact that
the energy cost to match independent amorphous configurations vanishes
for high temperatures seems to support the existence of the spinodal
crossover proposed in ~\cite{self:jstatmech09, self:Cammarota09b,
  self:Cammarota10}.

At the three lowest temperatures the excess energy seems to be well described 
by an exponential decay with $d$,
\begin{equation}
\Delta E (d) \sim e^{-d/\xi_E},
\label{eq:expdecDE}
\end{equation}
(see inset of Fig.~\ref{fig:DEvsT}).  We must note that, at variance
with the case of the point-to-set correlation, in the excess energy we
do not find any hint of non-exponentiality.  Moreover, the decay of
$\Delta E(d)$ defines a new lengthscale $l$.  We shall investigate in
the next Section whether $l$ can be identified with the point-to-set
correlation length $\xi$ or with the penetration length $\lambda$.

\subsection{Aspect ratio scaling}

It is interesting, and potentially useful, to analyze the excess
energy using some simple scaling relations, partially inspired by the
aspect-ratio-scaling technique introduced in
ref.~\cite{tension:Carter02}.  The basic ideas of this subsection have
been already used in the naive RFOT argument of the previous section.

The relevant lengthscales for $\Delta E$ are $d, l$ and $L$, the
longitudinal size of the sandwich.  The first thing we can say is that
the excess energy will scale like a length to the exponent
$\theta$ (which is basically a definition of the stiffness exponent).
Hence 
\begin{equation}
\Delta E \sim Y L^\theta f(d/L, l/L)  ,
\end{equation}
where $Y$ is the (generalized) surface tension.  One can choose any of
the three lengths to fix the dimensions by appropriately changing the
scaling function $f$.  The second requirement is that the energy and
the excess energy must be extensive: in the limit $L\gg d$, the
$\Delta E$ from different pieces of the surface must add up.  This
implies that,
\begin{equation}
\Delta E \sim L^2   ,
\end{equation}
a relation very well obeyed by our data.  For this to be true we need that
\begin{equation}
f(d/L,d/l) \sim (L/d)^{2-\theta} g(d/l)  .
\end{equation}
Moreover, as we have seen from the data, the scale $l$ seems well
set by an exponential decay, so it is reasonable to assume $g(x)=e^{-x}$,
so that
\begin{equation}
\Delta E (d) \sim Y L^2\; \frac{1}{d^{2-\theta}} e^{-d/l}  .
\label{arse}
\end{equation}
This is an interesting formula, and one could in principle use it to
fit the stiffness exponent $\theta$. In particular, the formula
suggests that, if a purely exponential fit is satisfactory (as in our
case), then $\theta\sim 2$.  In practice, the formula is useful to
discriminate different values of $\theta$ only for large $d$; but for
$\Delta E$ to be nonzero at large $d$, we need very large values of
$l$, i.e.\ very low temperatures, which we do not have.  In fact, any
exponent $\theta$ in the interval $[1,2]$ does an equally good job in
fitting our data for $\Delta E(d)$.  In particular, distinguishing
between $\theta=3/2$ and $\theta=2$ is completely out of the question.
Yet, the method is conceptually interesting, and future simulations,
at lower $T$, may eventually use it to determine the stiffness
exponent.

\section{Comparison of the different lengthscales and the Ising case}
\label{sec:lengthsc-comp-ising}

Let us summarize the three lengthscales we have measured.  The first
one is the point-to-set correlation length, $\xi$, defined as the
decay scale of the overlap measured at the centre of a sandwich of
half-width $d$,
\begin{equation}
q_c(d) \sim \exp[-(d/\xi)^\zeta] + q_0  .
\label{qui}
\end{equation}
Previous investigations suggest that this is the true static
correlation length of the system, the one relevant for the structural
rearrangement \cite{self:jcp12}.  Moreover, there is evidence
\cite{self:nphys08} that $\xi$ has to be identified with the RFOT
correlation length, discussed above.  The remarkable feature of the
point-to-set correlation length is that its associated correlation
function has a non-exponential decay at low temperature.

The second lengthscale is the penetration length $\lambda$, which regulates
the decay of the overlap off a single wall in a semi-infinite geometry, 
\begin{equation}
q(z) \sim \exp[-z/\lambda] +q_0 .
\label{quo}
\end{equation}
This lengthscale seems to have a different physical meaning than the 
correlation length $\xi$, as also suggested in \cite{pinning:cammarota12}.
It seems to embody the extent to which the effect of a single frozen wall
penetrates into the system, rather than the average size of a rearranging region.
At variance with the point-to-set correlation length, $\lambda$ regulates
a purely exponential decay of the overlap.

Finally, we measured the lengthscale $l$ associated to the decay in $d$ of
the excess energy produced by imposing ``anti-parallel'' boundary conditions, 
\begin{equation}
\Delta E(d) = E_{\alpha\beta}-E_{\alpha\alpha} \sim  \exp[-d/l]   .
\label{qua}
\end{equation}
As in the case of the penetration length, the excess energy lengthscale 
$l$ is associated to a purely exponential decay, at least down to our lowest
available temperature.

What can be said about the quantitative relationship (if any) between
these three lengthscales? We report them all in
Table~\ref{tab:lengths}, together with the anomaly exponent $\zeta$.

\begin{table}[htdp]
\label{tab:lengths}
  \caption{Point-to-set correlation length $\xi$, and anomaly
    exponent $\zeta$, from a fit of Eq.~\ref{qui}; penetration length
    $\lambda$, from a fit of Eq.~\ref{quo}; and excess energy decay
    lengthscale $l$, from a fit of Eq.~\ref{qua}. At the highest
    temperature the value of $l$ has large uncertainty as we have very
    few nonzero values of $\Delta E$.}
\begin{center}
\begin{tabular}{c|c|c|c|c}
$T$ & $\xi $ & $\zeta$ & $\lambda$ & $l$ \\
\hline 
$0.482$ & 0.48 & 1 & 0.47 & 0.15   \\
$0.350$ & 0.56 &  1 & 0.56 & 0.33   \\
$0.246$ & 1.50 & 2.1 & 0.69 & 0.43  \\
$0.202$ & 2.00 & 2.7 & 0.79 &  0.50 \\
\end{tabular}
\end{center}
\end{table}

One could object that much of the comparison depends on the fitting
procedure of the data, which is not a nice thing.  This is certainly a
concern.  However, we notice that extracting the lengthscales by
crossing the various functions with arbitrary threshold would not be
any better, for two reasons: first, in presence of a nonexponential
decay (as $q_c(d)$ unmistakably has), with a $T$-dependent anomalous
exponent $\zeta$, the arbitrary value of the threshold can strongly
bias the dependence of $\xi$ on $T$; secondly, these are dimensionally
different, inhomogeneous functions, so it would be hard to choose
coherently a crossing point for each of them.  An honest fit is the
best we can do.

\begin{figure}
  \includegraphics[width=\columnwidth]{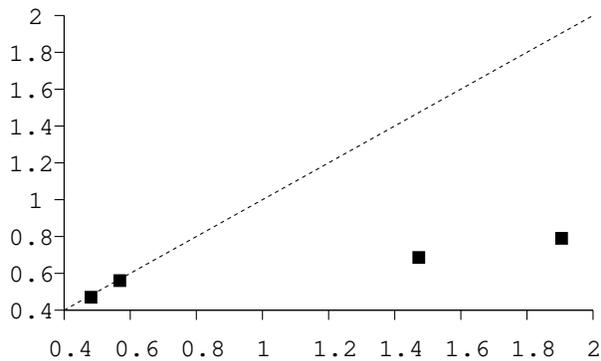}
  \caption{Penetration length $\lambda$ vs.\ point-to-set correlation
    length $\xi$.}
  \label{fig:xivslambda}
\end{figure}

\begin{figure}
  \includegraphics[width=\columnwidth]{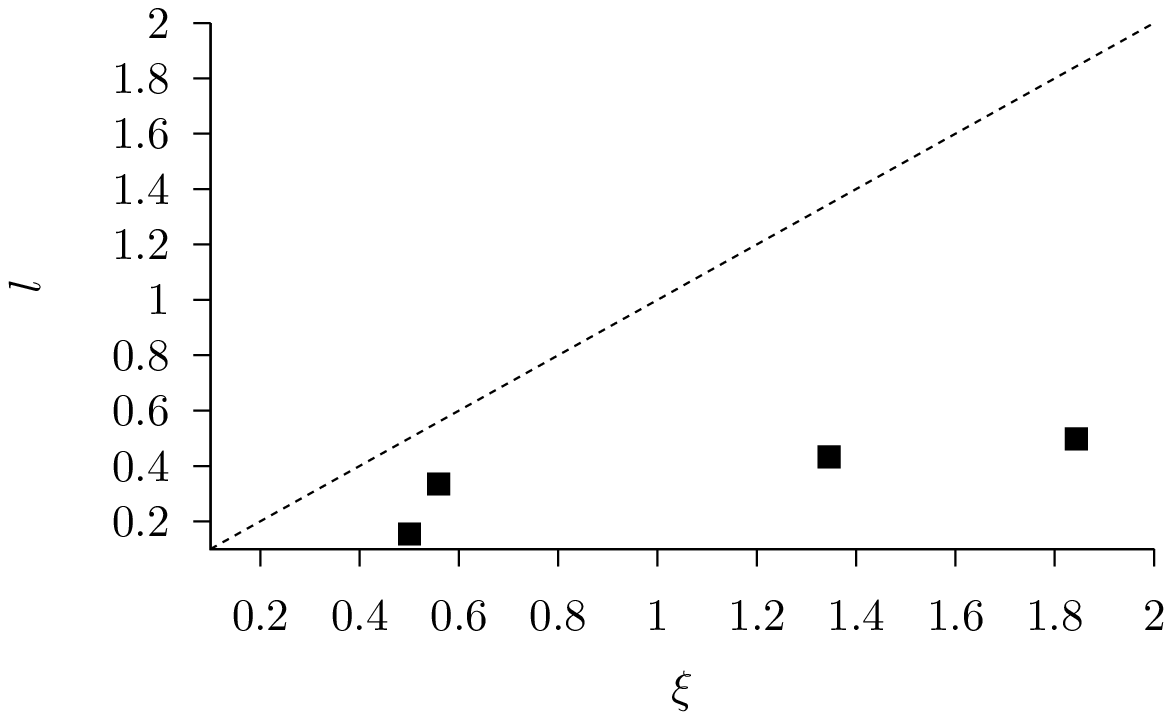}
  \caption{Energy decay length $l$ vs.\ point-to-set correlation
    length $\xi$.}
  \label{fig:xivsl}
\end{figure}

From the table we see that the correlation length $\xi$ is larger than
the other two scales.  Of course, what really matters is their mutual
$T$-dependence, namely: is there any of them that grows significantly
faster than the others, when lowering $T$?  More precisely, we would
like to understand whether or not these lengths are ruled by different
exponents.  Because of this, comparing the three plots, $\xi(T),
\lambda(T), l(T)$, is not a good idea: constant factors would show up
as increasing differences, conveying the (wrong) idea that one length
is growing faster than the other.  The best thing to do is to plot one
lengthscale vs.\ the other, parametrically in $T$.  This is what we do
in Figs.~\ref{fig:xivslambda} and~\ref{fig:xivsl}.

The result of this comparison is unfortunately not conclusive.  Even
though, as we already said, the correlation length $\xi$ is larger
than the other two, all mutual dependences are not far from
linear. This means that, with such data, we cannot claim that $\xi$ is
growing with an exponent significantly different from the other two,
which would be the only proof of a qualitative difference between
these scales.  Of course, our data do not either rule this out.

\begin{figure}
  \includegraphics[width=\columnwidth]{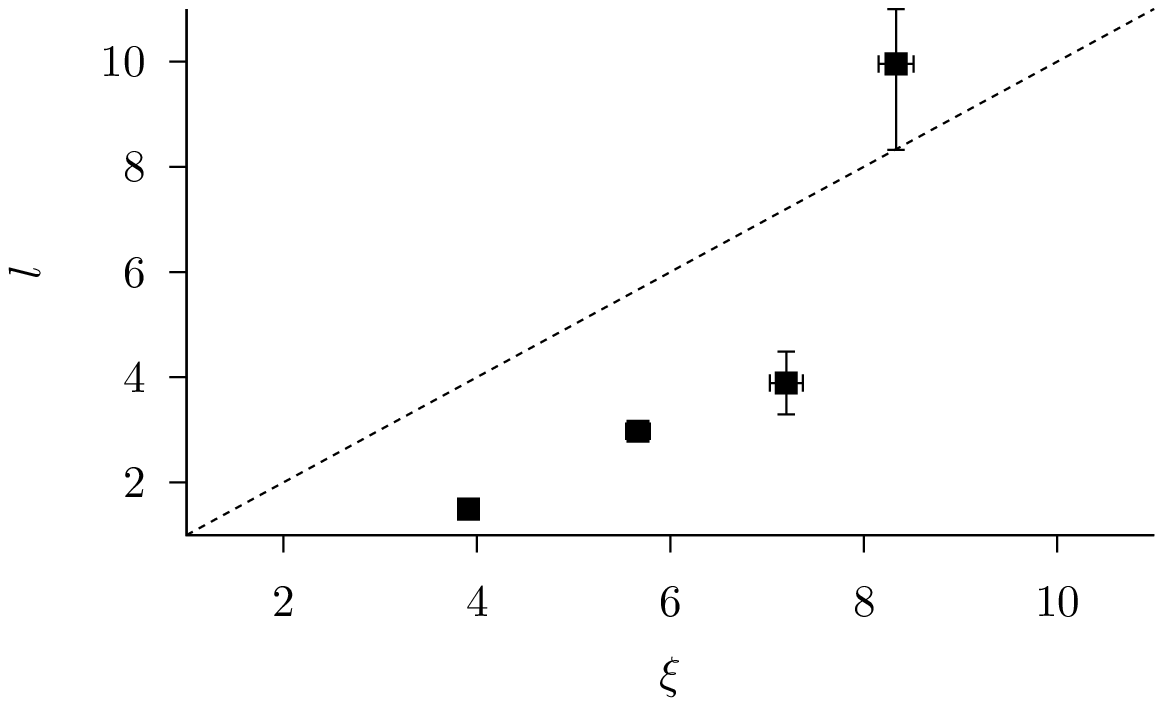}
  \caption{Energy decay length vs.\ two-spin correlation length for
    the Ising model in the square lattice.  Data are from Monte Carlo
    simulations on a $100\times 100$ lattice with single-flip
    Metropolis dynamics performed above the critical point, at
    temperatures $T=2.5J$, $T=2.4J$, $T=2.35J$ and $T=2.32J$, where
    $J$ is the Ising coupling constant (the critical point is
    $T_c\approx 2.269J$). The correlation length was obtained from a
    fit of the spin-spin space correlation function $C(r)=\langle
    S(0)S(r)\rangle$.  To determine the length $l$, sandwich
    configurations were prepared as explained for the liquid case,
    measuring the excess energy $\Delta
    E(d)=E_{\alpha\beta}-E_{\alpha\alpha}$ for $d=1$, 2, 3, 4, 5, 7.5,
    10, 15, 20, 25 and 45 lattice spacings, and fitting to an
    exponential decay.  }
  \label{fig:xivsl-ising}
\end{figure}

In such a murky situation, some conceptual help may perhaps come from
the well-known Ising model.  There one can use (true) anti-parallel
boundary conditions to force a domain wall (below the critical
temperature $Tc$).  Then there is a finite surface tension and the
excess free energy (anti-parallel minus parallel) in the limit
$d\to\infty$ tends to the finite value $\sigma L^{d-1}$, where
$\sigma$ is the surface tension.  For $T>T_c$, one can instead impose
(similarly to what we have done above) two {\it different}
paramagnetic configurations on the two sides of the sandwich and
measure the excess energy.  In this case we expect the excess energy
to decay to zero for large $d$, but on what scale does this happen?
Simulations in two-dimensions show that the excess energy decays
exponentially with a lengthscale $l$, which we can compare with the
Ising correlation length $\xi$, calculated from the standard spin-spin
space correlation function.  Data show that $l$ and $\xi$ scale
linearly with each other (parametrically in $T$,
Fig.~\ref{fig:xivsl-ising}), though $l$ is somewhat $2$ smaller than
$\xi$.  Therefore, in Ising, correlation decay and excess energy decay
seem to track each other quite closely.

\section{Conclusions}
\label{final} 

Similarly to what we found in the spherical geometry \cite{self:nphys08}, the sandwich 
data show a crossover from exponential to nonexponential relaxation of the 
point-to-set correlation function, upon lowering the temperature.
We remark that such crossover is one of the very few (if not the only one)
{\it static} landmarks differentiating at the qualitative level the fluid phase from the deeply 
supercooled phase in glass-forming liquids. Having found this feature now in 
two different geometries makes it quite a robust phenomenon.

Up to now, the only reasonable explanation of such sharper-than-exponential
relaxation has been given in the context of RFOT. We reported a naive RFOT
argument for the sandwich and showed that there should be no essential 
variations (at least in $3$ dimensions) with respect to the standard argument
one uses in the cavity geometry. Our sandwich results therefore give further support
to the theoretical connection between point-to-set nonexponential relaxation and RFOT.

A more thorough formulation of the RFOT argument, based on a renormalization 
group framework, could turn the nonexponential, but smooth, drop of the overlap 
at $d=d_\mathrm{RFOT}$, into a true transition, in the limit $L\to\infty$ 
\cite{pinning:cammarota12}. The very existence of such limit would be 
one of the main benefits of the sandwich vs. the cavity geometry.
However, this transition is supposed to be smeared out by disorder (surface 
tension fluctuations) in a three-dimensional sandwich \cite{pinning:cammarota12},
and indeed, by substantially increasing $L$, we do not find any relevant change in 
the point-to-set correlation function. 

A different experiment consists in measuring the overlap decay off a
single all.  In this case, we found a behaviour rather different from
the point-to-set correlation function. First, and most important, this
decay is purely exponential, even a the lowest temperature studied,
where the point-to-set correlation function is clearly nonexponential.
Hence, the single wall seems less than ideal to characterize at the
qualitative level the deeply supercooled phase.

Second, the lengthscale of this single-wall decay, i.e.\ the
penetration length $\lambda$, seems to be smaller than the
point-to-set correlation length, $\xi$. According to
ref.~\cite{mosaic:cammarota11}, both $\xi$ and $\lambda$ should
diverge at $T_k$, but with different exponents, in particular the
divergence of $\xi$ should be sharper than that of $\lambda$.  This is
due to the fact that $\lambda$ is not controlled by the RFOT entropy
vs.\ surface tension competition mechanism so directly as $\xi$ is.
Even though we are far (to say the least) from the $T\sim T_k$ region
where the RG arguments of ref.~\cite{mosaic:cammarota11} hold, we can
at least say that our numerical data are not in contradiction with
this scenario.

We used ``anti-parallel'' boundary conditions in the sandwich to
measure the excess energy associated to an interface. This quantity
seems to decay purely exponentially with the half-width $d$ of the
sandwich, over a lengthscale $l$ that grows by lowering the
temperature. Unfortunately, the value of $l$ we obtain even at the
lowest $T$ is not large enough to make it possible an estimate of the
stiffness exponent $\theta$ using aspect ratio scaling.  However, the
technique seems promising in this context, and perhaps future
simulation will reach a regime able to discriminate between different
(theoretical) values of $\theta$.

An obvious question is whether and how the lengthscale of the excess
energy $l$ is related to the other two lengthscales, and in particular
to the correlation length, $\xi$. We do not have a final answer to
this question. Even though $\xi$ seems to be quantitatively larger
than $l$ (about a factor 4), there is no clear evidence of a nonlinear
connection between the two lengths. On one hand, by following an
economy criterion, we are tempted to conclude that point-to-set and
excess energy are regulated by {\it one} lengthscale, as it happens in
the Ising model above $T_c$. On the other hand, the very different
kind of relaxation (nonexponential for the point-to-set correlation
function, exponential for the excess energy), and the fact that $l$ is
significantly smaller than $\xi$, seem to suggest otherwise.  We
cannot but leave the question open.

Finally, the Ising example calls for caution in the interpretation of
our results about the excess energy.  The fact that in a purely
paramagnetic state one finds a behaviour of $\Delta E$ vs. $d$ so
similar to the glass-forming case, means that a finite $\Delta E$ is
not by itself proof of the existence of a surface tension.  This calls
for a through investigation of the possible {\it entropic}
contribution to the glassy surface tension (which we do not measure
here).

\begin{acknowledgments} We warmly thank Giorgio Parisi for making us
  familiar with the aspect ratio scaling technique.  We also thank
  Giulio Biroli and Chiara Cammarota for several important
  discussions.  TSG thanks the Dipartimento di Fisica of the
  \emph{Sapienza} Universit\'a di Roma and ISC (CNR, Rome) for
  hospitality.  The work of GG is supported by the ``Granular-Chaos''
  project, funded by Italian MIUR under the grant number RBID08Z9JE.
  PV was partly supported by MICINN (Spain) through Research Contract
  Nos.\ FIS2009-12648-C03-01 and FIS2008-01323 (PV).  TSG was
  partially supported by ANPCyT (Argentina).
\end{acknowledgments}

\bibliography{references}

\begin{thebibliography}{10}%
\makeatletter
\providecommand \@ifxundefined [1]{%
 \ifx #1\undefined \expandafter \@firstoftwo
 \else \expandafter \@secondoftwo
\fi
}%
\providecommand \@ifnum [1]{%
 \ifnum #1\expandafter \@firstoftwo
 \else \expandafter \@secondoftwo
\fi
}%
\providecommand \enquote [1]{``#1''}%
\providecommand \bibnamefont  [1]{#1}%
\providecommand \bibfnamefont [1]{#1}%
\providecommand \citenamefont [1]{#1}%
\providecommand\href[0]{\@sanitize\@href}%
\providecommand\@href[1]{\endgroup\@@startlink{#1}\endgroup\@@href}%
\providecommand\@@href[1]{#1\@@endlink}%
\providecommand \@sanitize [0]{\begingroup\catcode`\&12\catcode`\#12\relax}%
\@ifxundefined \pdfoutput {\@firstoftwo}{%
 \@ifnum{\z@=\pdfoutput}{\@firstoftwo}{\@secondoftwo}%
}{%
 \providecommand\@@startlink[1]{\leavevmode\special{html:<a href="#1">}}%
 \providecommand\@@endlink[0]{\special{html:</a>}}%
}{%
 \providecommand\@@startlink[1]{%
  \leavevmode
  \pdfstartlink
   attr{/Border[0 0 1 ]/H/I/C[0 1 1]}%
   user{/Subtype/Link/A<</Type/Action/S/URI/URI(#1)>>}%
  \relax
 }%
 \providecommand\@@endlink[0]{\pdfendlink}%
}%
\providecommand \url  [0]{\begingroup\@sanitize \@url }%
\providecommand \@url [1]{\endgroup\@href {#1}{\urlprefix}}%
\providecommand \urlprefix [0]{URL }%
\providecommand \Eprint[0]{\href }%
\@ifxundefined \urlstyle {%
  \providecommand \doi [1]{doi:\discretionary{}{}{}#1}%
}{%
  \providecommand \doi [0]{doi:\discretionary{}{}{}\begingroup
  \urlstyle{rm}\Url }%
}%
\providecommand \doibase [0]{http://dx.doi.org/}%
\providecommand \Doi[1]{\href{\doibase#1}}%
\providecommand \bibAnnote [3]{%
  \BibitemShut{#1}%
  \begin{quotation}\noindent
    \textsc{Key:}\ #2\\\textsc{Annotation:}\ #3%
  \end{quotation}%
}%
\providecommand \bibAnnoteFile [2]{%
  \IfFileExists{#2}{\bibAnnote {#1} {#2} {\input{#2}}}{}%
}%
\providecommand \typeout [0]{\immediate \write \m@ne }%
\providecommand \selectlanguage [0]{\@gobble}%
\providecommand \bibinfo [0]{\@secondoftwo}%
\providecommand \bibfield [0]{\@secondoftwo}%
\providecommand \translation [1]{[#1]}%
\providecommand \BibitemOpen[0]{}%
\providecommand \bibitemStop [0]{}%
\providecommand \bibitemNoStop [0]{.\EOS\space}%
\providecommand \EOS [0]{\spacefactor3000\relax}%
\providecommand \BibitemShut [1]{\csname bibitem#1\endcsname}%
\bibitem{heterogeneities:sillescu99}%
  \BibitemOpen
  \bibfield{author}{%
  \bibinfo {author} {\bibfnamefont{H.}~\bibnamefont{Sillescu}},\ }%
  \bibfield{journal}{%
  \Doi{DOI: 10.1016/S0022-3093(98)00831-X}{\bibinfo {journal} {J. Non-Cryst.
  Solids}}\ }%
  \textbf{\bibinfo {volume} {243}},\ \bibinfo {pages} {81 } (\bibinfo {year}
  {1999}),\ ISSN \bibinfo {issn} {0022-3093}%
  \bibAnnoteFile{NoStop}{heterogeneities:sillescu99}%
\bibitem{review:Ediger00}%
  \BibitemOpen
  \bibfield{author}{%
  \bibinfo {author} {\bibfnamefont{M.~D.}\ \bibnamefont{Ediger}},\ }%
  \bibfield{journal}{%
  \bibinfo {journal} {Annu. Rev. Phys. Chem.}\ }%
  \textbf{\bibinfo {volume} {51}},\ \bibinfo {pages} {99} (\bibinfo {year}
  {2000})%
  \bibAnnoteFile{NoStop}{review:Ediger00}%
\bibitem{ISRAELOFF00}%
  \BibitemOpen
  \bibfield{author}{%
  \bibinfo {author} {\bibfnamefont{E.}~\bibnamefont{Vidal-Russell}}\ and\
  \bibinfo {author} {\bibfnamefont{N.~E.}\ \bibnamefont{Israeloff}},\ }%
  \bibfield{journal}{%
  \bibinfo {journal} {Nature}\ }%
  \textbf{\bibinfo {volume} {408}},\ \bibinfo {pages} {695} (\bibinfo {year}
  {2000})%
  \bibAnnoteFile{NoStop}{ISRAELOFF00}%
\bibitem{berthier03b}%
  \BibitemOpen
  \bibfield{author}{%
  \bibinfo {author} {\bibfnamefont{L.}~\bibnamefont{Berthier}}\ and\ \bibinfo
  {author} {\bibfnamefont{J.~P.}\ \bibnamefont{Garrahan}},\ }%
  \bibfield{journal}{%
  \bibinfo {journal} {J. Chem. Phys.}\ }%
  \textbf{\bibinfo {volume} {119}},\ \bibinfo {pages} {4367} (\bibinfo {year}
  {2003})%
  \bibAnnoteFile{NoStop}{berthier03b}%
\bibitem{Cugliandolo03}%
  \BibitemOpen
  \bibfield{author}{%
  \bibinfo {author} {\bibfnamefont{H.~E.}\ \bibnamefont{Castillo}}, \bibinfo
  {author} {\bibfnamefont{C.}~\bibnamefont{Chamon}}, \bibinfo {author}
  {\bibfnamefont{L.~F.}\ \bibnamefont{Cugliandolo}}, \bibinfo {author}
  {\bibfnamefont{J.~L.}\ \bibnamefont{Iguain}},\ and\ \bibinfo {author}
  {\bibfnamefont{M.~P.}\ \bibnamefont{Kennett}},\ }%
  \bibfield{journal}{%
  \bibinfo {journal} {Phys. Rev. B}\ }%
  \textbf{\bibinfo {volume} {68}},\ \bibinfo {pages} {134442} (\bibinfo {year}
  {2003})%
  \bibAnnoteFile{NoStop}{Cugliandolo03}%
\bibitem{BERTHIER03}%
  \BibitemOpen
  \bibfield{author}{%
  \bibinfo {author} {\bibfnamefont{L.}~\bibnamefont{Berthier}},\ }%
  \bibfield{journal}{%
  \bibinfo {journal} {Phys. Rev. Lett.}\ }%
  \textbf{\bibinfo {volume} {91}},\ \bibinfo {pages} {055701} (\bibinfo {year}
  {2003})%
  \bibAnnoteFile{NoStop}{BERTHIER03}%
\bibitem{BERTHIER04}%
  \BibitemOpen
  \bibfield{author}{%
  \bibinfo {author} {\bibfnamefont{L.}~\bibnamefont{Berthier}},\ }%
  \bibfield{journal}{%
  \bibinfo {journal} {Phys. Rev. E}\ }%
  \textbf{\bibinfo {volume} {69}},\ \bibinfo {pages} {020201} (\bibinfo {year}
  {2004})%
  \bibAnnoteFile{NoStop}{BERTHIER04}%
\bibitem{heterogeneities:sillescu02}%
  \BibitemOpen
  \bibfield{author}{%
  \bibinfo {author} {\bibfnamefont{H.}~\bibnamefont{Sillescu}}, \bibinfo
  {author} {\bibfnamefont{R.}~\bibnamefont{B\"ohmer}}, \bibinfo {author}
  {\bibfnamefont{G.}~\bibnamefont{Diezemann}},\ and\ \bibinfo {author}
  {\bibfnamefont{G.}~\bibnamefont{Hinze}},\ }%
  \bibfield{journal}{%
  \Doi{DOI: 10.1016/S0022-3093(02)01435-7}{\bibinfo {journal} {J. Non-Cryst.
  Solids}}\ }%
  \textbf{\bibinfo {volume} {307-310}},\ \bibinfo {pages} {16 } (\bibinfo
  {year} {2002}),\ ISSN \bibinfo {issn} {0022-3093}%
  \bibAnnoteFile{NoStop}{heterogeneities:sillescu02}%
\bibitem{heterogeneities:garrahan02}%
  \BibitemOpen
  \bibfield{author}{%
  \bibinfo {author} {\bibfnamefont{J.~P.}\ \bibnamefont{Garrahan}}\ and\
  \bibinfo {author} {\bibfnamefont{D.}~\bibnamefont{Chandler}},\ }%
  \bibfield{journal}{%
  \Doi{10.1103/PhysRevLett.89.035704}{\bibinfo {journal} {Phys. Rev. Lett.}}\
  }%
  \textbf{\bibinfo {volume} {89}},\ \bibinfo {pages} {035704} (\bibinfo {month}
  {Jul}\ \bibinfo {year} {2002})%
  \bibAnnoteFile{NoStop}{heterogeneities:garrahan02}%
\bibitem{heterogeneities:Berthier05}%
  \BibitemOpen
  \bibfield{author}{%
  \bibinfo {author} {\bibfnamefont{L.}~\bibnamefont{Berthier}}, \bibinfo
  {author} {\bibfnamefont{G.}~\bibnamefont{Biroli}}, \bibinfo {author}
  {\bibfnamefont{J.-P.}\ \bibnamefont{Bouchaud}}, \bibinfo {author}
  {\bibfnamefont{L.}~\bibnamefont{Cipelletti}}, \bibinfo {author}
  {\bibfnamefont{D.~E.}\ \bibnamefont{Masri}}, \bibinfo {author}
  {\bibfnamefont{D.}~\bibnamefont{L'H{\^o}te}}, \bibinfo {author}
  {\bibfnamefont{F.}~\bibnamefont{Ladieu}},\ and\ \bibinfo {author}
  {\bibfnamefont{M.}~\bibnamefont{Pierno}},\ }%
  \bibfield{journal}{%
  \bibinfo {journal} {Science}\ }%
  \textbf{\bibinfo {volume} {310}},\ \bibinfo {pages} {1797} (\bibinfo {year}
  {2005})%
  \bibAnnoteFile{NoStop}{heterogeneities:Berthier05}%
\bibitem{heterogeneities:berthier07a}%
  \BibitemOpen
  \bibfield{author}{%
  \bibinfo {author} {\bibfnamefont{L.}~\bibnamefont{Berthier}}, \bibinfo
  {author} {\bibfnamefont{G.}~\bibnamefont{Biroli}}, \bibinfo {author}
  {\bibfnamefont{J.-P.}\ \bibnamefont{Bouchaud}}, \bibinfo {author}
  {\bibfnamefont{W.}~\bibnamefont{Kob}}, \bibinfo {author}
  {\bibfnamefont{K.}~\bibnamefont{Miyazaki}},\ and\ \bibinfo {author}
  {\bibfnamefont{D.~R.}\ \bibnamefont{Reichman}},\ }%
  \bibfield{journal}{%
  \Doi{10.1063/1.2721554}{\bibinfo {journal} {J. Chem. Phys.}}\ }%
  \textbf{\bibinfo {volume} {126}},\ \bibinfo {eid} {184503} (\bibinfo {year}
  {2007})%
  \bibAnnoteFile{NoStop}{heterogeneities:berthier07a}%
\bibitem{heterogeneities:berthier07b}%
  \BibitemOpen
  \bibfield{author}{%
  \bibinfo {author} {\bibfnamefont{L.}~\bibnamefont{Berthier}}, \bibinfo
  {author} {\bibfnamefont{G.}~\bibnamefont{Biroli}}, \bibinfo {author}
  {\bibfnamefont{J.-P.}\ \bibnamefont{Bouchaud}}, \bibinfo {author}
  {\bibfnamefont{W.}~\bibnamefont{Kob}}, \bibinfo {author}
  {\bibfnamefont{K.}~\bibnamefont{Miyazaki}},\ and\ \bibinfo {author}
  {\bibfnamefont{D.~R.}\ \bibnamefont{Reichman}},\ }%
  \bibfield{journal}{%
  \Doi{10.1063/1.2721555}{\bibinfo {journal} {J. Chem. Phys.}}\ }%
  \textbf{\bibinfo {volume} {126}},\ \bibinfo {eid} {184504} (\bibinfo {year}
  {2007})%
  \bibAnnoteFile{NoStop}{heterogeneities:berthier07b}%
\bibitem{review:kivelson97}%
  \BibitemOpen
  \bibfield{author}{%
  \bibinfo {author} {\bibfnamefont{D.}~\bibnamefont{{Kivelson}}}, \bibinfo
  {author} {\bibfnamefont{G.}~\bibnamefont{{Tarjus}}},\ and\ \bibinfo {author}
  {\bibfnamefont{S.~A.}\ \bibnamefont{{Kivelson}}},\ }%
  \bibfield{journal}{%
  \bibinfo {journal} {Progr. Theor. Phys. Supp.}\ }%
  \textbf{\bibinfo {volume} {126}},\ \bibinfo {pages} {289} (\bibinfo {year}
  {1997})%
  \bibAnnoteFile{NoStop}{review:kivelson97}%
\bibitem{self:prl07}%
  \BibitemOpen
  \bibfield{author}{%
  \bibinfo {author} {\bibfnamefont{A.}~\bibnamefont{Cavagna}}, \bibinfo
  {author} {\bibfnamefont{T.~S.}\ \bibnamefont{Grigera}},\ and\ \bibinfo
  {author} {\bibfnamefont{P.}~\bibnamefont{Verrocchio}},\ }%
  \bibfield{journal}{%
  \Doi{10.1103/PhysRevLett.98.187801}{\bibinfo {journal} {Phys. Rev. Lett.}}\
  }%
  \textbf{\bibinfo {volume} {98}},\ \bibinfo {eid} {187801} (\bibinfo {year}
  {2007})%
  \bibAnnoteFile{NoStop}{self:prl07}%
\bibitem{landscape:widmer-cooper08}%
  \BibitemOpen
  \bibfield{author}{%
  \bibinfo {author} {\bibfnamefont{A.}~\bibnamefont{Widmer-Cooper}}, \bibinfo
  {author} {\bibfnamefont{H.}~\bibnamefont{Perry}}, \bibinfo {author}
  {\bibfnamefont{P.}~\bibnamefont{Harrowell}},\ and\ \bibinfo {author}
  {\bibfnamefont{D.~R.}\ \bibnamefont{Reichman}},\ }%
  \bibfield{journal}{%
  \bibinfo {journal} {Nature Phys.}\ }%
  \textbf{\bibinfo {volume} {4}},\ \bibinfo {pages} {711} (\bibinfo {year}
  {2008})%
  \bibAnnoteFile{NoStop}{landscape:widmer-cooper08}%
\bibitem{glassthermo:tanaka10}%
  \BibitemOpen
  \bibfield{author}{%
  \bibinfo {author} {\bibfnamefont{H.}~\bibnamefont{Tanaka}}, \bibinfo {author}
  {\bibfnamefont{T.}~\bibnamefont{Kawasaki}}, \bibinfo {author}
  {\bibfnamefont{H.}~\bibnamefont{Shintani}},\ and\ \bibinfo {author}
  {\bibfnamefont{K.}~\bibnamefont{Watanabe}},\ }%
  \bibfield{journal}{%
  \bibinfo {journal} {Nature Mater}\ }%
  \textbf{\bibinfo {volume} {9}},\ \bibinfo {pages} {324} (\bibinfo {year}
  {2010})%
  \bibAnnoteFile{NoStop}{glassthermo:tanaka10}%
\bibitem{glassthermo:coslovich2011}%
  \BibitemOpen
  \bibfield{author}{%
  \bibinfo {author} {\bibfnamefont{D.}~\bibnamefont{Coslovich}},\ }%
  \bibfield{journal}{%
  \Doi{10.1103/PhysRevE.83.051505}{\bibinfo {journal} {Phys. Rev. E}}\ }%
  \textbf{\bibinfo {volume} {83}},\ \bibinfo {pages} {051505} (\bibinfo {month}
  {May}\ \bibinfo {year} {2011})%
  \bibAnnoteFile{NoStop}{glassthermo:coslovich2011}%
\bibitem{Scheidler02}%
  \BibitemOpen
  \bibfield{author}{%
  \bibinfo {author} {\bibfnamefont{P.}~\bibnamefont{Scheidler}}, \bibinfo
  {author} {\bibfnamefont{W.}~\bibnamefont{Kob}}, \bibinfo {author}
  {\bibfnamefont{K.}~\bibnamefont{Binder}},\ and\ \bibinfo {author}
  {\bibfnamefont{G.}~\bibnamefont{Parisi}},\ }%
  \bibfield{journal}{%
  \bibinfo {journal} {Phil. Mag. B}\ }%
  \textbf{\bibinfo {volume} {82}},\ \bibinfo {pages} {283} (\bibinfo {year}
  {2002})%
  \bibAnnoteFile{NoStop}{Scheidler02}%
\bibitem{self:nphys08}%
  \BibitemOpen
  \bibfield{author}{%
  \bibinfo {author} {\bibfnamefont{G.}~\bibnamefont{Biroli}}, \bibinfo {author}
  {\bibfnamefont{J.-P.}\ \bibnamefont{Bouchaud}}, \bibinfo {author}
  {\bibfnamefont{A.}~\bibnamefont{Cavagna}}, \bibinfo {author}
  {\bibfnamefont{T.~S.}\ \bibnamefont{Grigera}},\ and\ \bibinfo {author}
  {\bibfnamefont{P.}~\bibnamefont{Verrocchio}},\ }%
  \bibfield{journal}{%
  \Doi{10.1038/nphys1050}{\bibinfo {journal} {Nature Phys.}}\ }%
  \textbf{\bibinfo {volume} {4}},\ \bibinfo {pages} {771} (\bibinfo {year}
  {2008})%
  \bibAnnoteFile{NoStop}{self:nphys08}%
\bibitem{confinement:berthier12}%
  \BibitemOpen
  \bibfield{author}{%
  \bibinfo {author} {\bibfnamefont{L.}~\bibnamefont{Berthier}}\ and\ \bibinfo
  {author} {\bibfnamefont{W.}~\bibnamefont{Kob}},\ }%
  \bibfield{journal}{%
  \Doi{10.1103/PhysRevE.85.011102}{\bibinfo {journal} {Phys. Rev. E}}\ }%
  \textbf{\bibinfo {volume} {85}},\ \bibinfo {pages} {011102} (\bibinfo {month}
  {Jan}\ \bibinfo {year} {2012})%
  \bibAnnoteFile{NoStop}{confinement:berthier12}%
\bibitem{confinement:kob12}%
  \BibitemOpen
  \bibfield{author}{%
  \bibinfo {author} {\bibfnamefont{W.}~\bibnamefont{Kob}}, \bibinfo {author}
  {\bibfnamefont{S.}~\bibnamefont{Rold\'an-Vargas}},\ and\ \bibinfo {author}
  {\bibfnamefont{L.}~\bibnamefont{Berthier}},\ }%
  \bibfield{journal}{%
  \Doi{http://dx.doi.org/10.1038/nphys2133}{\bibinfo {journal} {Nat. Phys.}}\
  }%
  \textbf{\bibinfo {volume} {8}},\ \bibinfo {pages} {164} (\bibinfo {year}
  {2012})%
  \bibAnnoteFile{NoStop}{confinement:kob12}%
\bibitem{correlation-length:hocky12}%
  \BibitemOpen
  \bibfield{author}{%
  \bibinfo {author} {\bibfnamefont{G.~M.}\ \bibnamefont{Hocky}}, \bibinfo
  {author} {\bibfnamefont{T.~E.}\ \bibnamefont{Markland}},\ and\ \bibinfo
  {author} {\bibfnamefont{D.~R.}\ \bibnamefont{Reichman}},\ }%
  \bibfield{journal}{%
  \Doi{10.1103/PhysRevLett.108.225506}{\bibinfo {journal} {Phys. Rev. Lett.}}\
  }%
  \textbf{\bibinfo {volume} {108}},\ \bibinfo {pages} {225506} (\bibinfo
  {month} {Jun}\ \bibinfo {year} {2012})%
  \bibAnnoteFile{NoStop}{correlation-length:hocky12}%
\bibitem{dynamics:montanari06}%
  \BibitemOpen
  \bibfield{author}{%
  \bibinfo {author} {\bibfnamefont{A.}~\bibnamefont{Montanari}}\ and\ \bibinfo
  {author} {\bibfnamefont{G.}~\bibnamefont{Semerjian}},\ }%
  \bibfield{journal}{%
  \Doi{10.1007/s10955-006-9175-y}{\bibinfo {journal} {J. Stat. Phys.}}\ }%
  \textbf{\bibinfo {volume} {125}},\ \bibinfo {pages} {23} (\bibinfo {year}
  {2006})%
  \bibAnnoteFile{NoStop}{dynamics:montanari06}%
\bibitem{mosaic:zarinelli10}%
  \BibitemOpen
  \bibfield{author}{%
  \bibinfo {author} {\bibfnamefont{E.}~\bibnamefont{Zarinelli}}\ and\ \bibinfo
  {author} {\bibfnamefont{S.}~\bibnamefont{Franz}},\ }%
  \bibfield{journal}{%
  \bibinfo {journal} {J. Stat. Mech.}\ }%
  \textbf{\bibinfo {volume} {2010}},\ \bibinfo {pages} {P04008} (\bibinfo
  {year} {2010})%
  \bibAnnoteFile{NoStop}{mosaic:zarinelli10}%
\bibitem{soft-spheres:bernu87}%
  \BibitemOpen
  \bibfield{author}{%
  \bibinfo {author} {\bibfnamefont{B.}~\bibnamefont{Bernu}}, \bibinfo {author}
  {\bibfnamefont{J.~P.}\ \bibnamefont{Hansen}}, \bibinfo {author}
  {\bibfnamefont{Y.}~\bibnamefont{Hiwatari}},\ and\ \bibinfo {author}
  {\bibfnamefont{G.}~\bibnamefont{Pastore}},\ }%
  \bibfield{journal}{%
  \Doi{10.1103/PhysRevA.36.4891}{\bibinfo {journal} {Phys. Rev. A}}\ }%
  \textbf{\bibinfo {volume} {36}},\ \bibinfo {pages} {4891} (\bibinfo {month}
  {Nov}\ \bibinfo {year} {1987})%
  \bibAnnoteFile{NoStop}{soft-spheres:bernu87}%
\bibitem{algorithm:Grigera01}%
  \BibitemOpen
  \bibfield{author}{%
  \bibinfo {author} {\bibfnamefont{T.~S.}\ \bibnamefont{Grigera}}\ and\
  \bibinfo {author} {\bibfnamefont{G.}~\bibnamefont{Parisi}},\ }%
  \bibfield{journal}{%
  \Doi{10.1103/PhysRevE.63.045102}{\bibinfo {journal} {Phys. Rev. E}}\ }%
  \textbf{\bibinfo {volume} {63}},\ \bibinfo {pages} {045102} (\bibinfo {month}
  {Mar}\ \bibinfo {year} {2001})%
  \bibAnnoteFile{NoStop}{algorithm:Grigera01}%
\bibitem{landscape:brumer04}%
  \BibitemOpen
  \bibfield{author}{%
  \bibinfo {author} {\bibfnamefont{Y.}~\bibnamefont{Brumer}}\ and\ \bibinfo
  {author} {\bibfnamefont{D.~R.}\ \bibnamefont{Reichman}},\ }%
  \bibfield{journal}{%
  \Doi{10.1103/PhysRevE.69.041202}{\bibinfo {journal} {Phys. Rev. E}}\ }%
  \textbf{\bibinfo {volume} {69}},\ \bibinfo {pages} {041202} (\bibinfo {month}
  {Apr}\ \bibinfo {year} {2004})%
  \bibAnnoteFile{NoStop}{landscape:brumer04}%
\bibitem{mosaic:kirkpatrick87}%
  \BibitemOpen
  \bibfield{author}{%
  \bibinfo {author} {\bibfnamefont{T.~R.}\ \bibnamefont{Kirkpatrick}}\ and\
  \bibinfo {author} {\bibfnamefont{P.~G.}\ \bibnamefont{Wolynes}},\ }%
  \bibfield{journal}{%
  \Doi{10.1103/PhysRevB.36.8552}{\bibinfo {journal} {Phys. Rev. B}}\ }%
  \textbf{\bibinfo {volume} {36}},\ \bibinfo {pages} {8552} (\bibinfo {month}
  {Dec}\ \bibinfo {year} {1987})%
  \bibAnnoteFile{NoStop}{mosaic:kirkpatrick87}%
\bibitem{mosaic:kirkpatrick89}%
  \BibitemOpen
  \bibfield{author}{%
  \bibinfo {author} {\bibfnamefont{T.~R.}\ \bibnamefont{Kirkpatrick}}, \bibinfo
  {author} {\bibfnamefont{D.}~\bibnamefont{Thirumalai}},\ and\ \bibinfo
  {author} {\bibfnamefont{P.~G.}\ \bibnamefont{Wolynes}},\ }%
  \bibfield{journal}{%
  \bibinfo {journal} {Phys. Rev. A}\ }%
  \textbf{\bibinfo {volume} {40}},\ \bibinfo {pages} {1045} (\bibinfo {year}
  {1989})%
  \bibAnnoteFile{NoStop}{mosaic:kirkpatrick89}%
\bibitem{mosaic:bouchaud04}%
  \BibitemOpen
  \bibfield{author}{%
  \bibinfo {author} {\bibfnamefont{J.-P.}\ \bibnamefont{Bouchaud}}\ and\
  \bibinfo {author} {\bibfnamefont{G.}~\bibnamefont{Biroli}},\ }%
  \bibfield{journal}{%
  \bibinfo {journal} {J. Chem. Phys.}\ }%
  \textbf{\bibinfo {volume} {121}},\ \bibinfo {pages} {7347} (\bibinfo {year}
  {2004})%
  \bibAnnoteFile{NoStop}{mosaic:bouchaud04}%
\bibitem{pinning:cammarota12}%
  \BibitemOpen
  \bibfield{author}{%
  \bibinfo {author} {\bibfnamefont{C.}~\bibnamefont{Cammarota}}\ and\ \bibinfo
  {author} {\bibfnamefont{G.}~\bibnamefont{{Biroli}}},\ }%
  \bibfield{journal}{%
  \bibinfo {journal} {Proc. Natl. Acad. Sci. USA}\ }%
  \textbf{\bibinfo {volume} {109}},\ \bibinfo {pages} {8850} (\bibinfo {month}
  {May}\ \bibinfo {year} {2012})%
  \bibAnnoteFile{NoStop}{pinning:cammarota12}%
\bibitem{glass:kurchan11}%
  \BibitemOpen
  \bibfield{author}{%
  \bibinfo {author} {\bibfnamefont{J.}~\bibnamefont{Kurchan}}\ and\ \bibinfo
  {author} {\bibfnamefont{D.}~\bibnamefont{Levine}},\ }%
  \bibfield{journal}{%
  \bibinfo {journal} {J. Phys. A: Math. Theor.}\ }%
  \textbf{\bibinfo {volume} {44}},\ \bibinfo {pages} {035001} (\bibinfo {year}
  {2011})%
  \bibAnnoteFile{NoStop}{glass:kurchan11}%
\bibitem{mosaic:cammarota12b}%
  \BibitemOpen
  \bibfield{author}{%
  \bibinfo {author} {\bibfnamefont{C.}~\bibnamefont{{Cammarota}}}\ and\
  \bibinfo {author} {\bibfnamefont{G.}~\bibnamefont{{Biroli}}},\ }%
  \bibfield{journal}{%
  \Doi{10.1209/0295-5075/98/36005}{\bibinfo {journal} {Europhys. Lett.}}\ }%
  \textbf{\bibinfo {volume} {98}},\ \bibinfo {pages} {36005} (\bibinfo {month}
  {May}\ \bibinfo {year} {2012})%
  \bibAnnoteFile{NoStop}{mosaic:cammarota12b}%
\bibitem{kirkpatrick89}%
  \BibitemOpen
  \bibfield{author}{%
  \bibinfo {author} {\bibfnamefont{T.}~\bibnamefont{Kirkpatrick}}, \bibinfo
  {author} {\bibfnamefont{D.}~\bibnamefont{Thirumalai}},\ and\ \bibinfo
  {author} {\bibfnamefont{P.}~\bibnamefont{Wolynes}},\ }%
  \bibfield{journal}{%
  \bibinfo {journal} {Phys. Rev. A}\ }%
  \textbf{\bibinfo {volume} {40}},\ \bibinfo {pages} {1045} (\bibinfo {year}
  {1989})%
  \bibAnnoteFile{NoStop}{kirkpatrick89}%
\bibitem{mosaic:franz11}%
  \BibitemOpen
  \bibfield{author}{%
  \bibinfo {author} {\bibfnamefont{S.}~\bibnamefont{{Franz}}}\ and\ \bibinfo
  {author} {\bibfnamefont{G.}~\bibnamefont{{Semerjian}}},\ }%
  \bibfield{journal}{%
  \bibinfo {journal} {ArXiv e-prints}}%
   (\bibinfo {month} {Sep.}\ \bibinfo {year} {2010}),\
  \Eprint{http://arxiv.org/abs/1009.5248}{arXiv:1009.5248
  [cond-mat.stat-mech]}%
  \bibAnnoteFile{NoStop}{mosaic:franz11}%
\bibitem{self:jstatmech09}%
  \BibitemOpen
  \bibfield{author}{%
  \bibinfo {author} {\bibfnamefont{C.}~\bibnamefont{Cammarota}}, \bibinfo
  {author} {\bibfnamefont{A.}~\bibnamefont{Cavagna}}, \bibinfo {author}
  {\bibfnamefont{G.}~\bibnamefont{Gradenigo}}, \bibinfo {author}
  {\bibfnamefont{T.~S.}\ \bibnamefont{Grigera}},\ and\ \bibinfo {author}
  {\bibfnamefont{P.}~\bibnamefont{Verrocchio}},\ }%
  \bibfield{journal}{%
  \bibinfo {journal} {J. Stat. Mech.}\ }%
  \textbf{\bibinfo {volume} {2009}},\ \bibinfo {pages} {L12002} (\bibinfo
  {year} {2009})%
  \bibAnnoteFile{NoStop}{self:jstatmech09}%
\bibitem{self:Cammarota09b}%
  \BibitemOpen
  \bibfield{author}{%
  \bibinfo {author} {\bibfnamefont{C.}~\bibnamefont{Cammarota}}, \bibinfo
  {author} {\bibfnamefont{A.}~\bibnamefont{Cavagna}}, \bibinfo {author}
  {\bibfnamefont{G.}~\bibnamefont{Gradenigo}}, \bibinfo {author}
  {\bibfnamefont{T.~S.}\ \bibnamefont{Grigera}},\ and\ \bibinfo {author}
  {\bibfnamefont{P.}~\bibnamefont{Verrocchio}},\ }%
  \bibfield{journal}{%
  \Doi{10.1063/1.3257739}{\bibinfo {journal} {J. Chem. Phys.}}\ }%
  \textbf{\bibinfo {volume} {131}},\ \bibinfo {eid} {194901} (\bibinfo {year}
  {2009})%
  \bibAnnoteFile{NoStop}{self:Cammarota09b}%
\bibitem{self:Cammarota10}%
  \BibitemOpen
  \bibfield{author}{%
  \bibinfo {author} {\bibfnamefont{C.}~\bibnamefont{Cammarota}}, \bibinfo
  {author} {\bibfnamefont{A.}~\bibnamefont{Cavagna}}, \bibinfo {author}
  {\bibfnamefont{I.}~\bibnamefont{Giardina}}, \bibinfo {author}
  {\bibfnamefont{G.}~\bibnamefont{Gradenigo}}, \bibinfo {author}
  {\bibfnamefont{T.~S.}\ \bibnamefont{Grigera}}, \bibinfo {author}
  {\bibfnamefont{G.}~\bibnamefont{Parisi}},\ and\ \bibinfo {author}
  {\bibfnamefont{P.}~\bibnamefont{Verrocchio}},\ }%
  \bibfield{journal}{%
  \Doi{10.1103/PhysRevLett.105.055703}{\bibinfo {journal} {Phys. Rev. Lett.}}\
  }%
  \textbf{\bibinfo {volume} {105}},\ \bibinfo {pages} {055703} (\bibinfo
  {month} {Jul}\ \bibinfo {year} {2010})%
  \bibAnnoteFile{NoStop}{self:Cammarota10}%
\bibitem{tension:Carter02}%
  \BibitemOpen
  \bibfield{author}{%
  \bibinfo {author} {\bibfnamefont{A.}~\bibnamefont{Carter}}, \bibinfo {author}
  {\bibfnamefont{A.}~\bibnamefont{Bray}},\ and\ \bibinfo {author}
  {\bibfnamefont{M.}~\bibnamefont{Moore}},\ }%
  \bibfield{journal}{%
  \Doi{10.1103/PhysRevLett.88.077201}{\bibinfo {journal} {Phys. Rev. Lett.}}\
  }%
  \textbf{\bibinfo {volume} {88}},\ \bibinfo {pages} {18} (\bibinfo {year}
  {2002}),\ ISSN \bibinfo {issn} {0031-9007}%
  \bibAnnoteFile{NoStop}{tension:Carter02}%
\bibitem{self:jcp12}%
  \BibitemOpen
  \bibfield{author}{%
  \bibinfo {author} {\bibfnamefont{A.}~\bibnamefont{Cavagna}}, \bibinfo
  {author} {\bibfnamefont{T.~S.}\ \bibnamefont{Grigera}},\ and\ \bibinfo
  {author} {\bibfnamefont{P.}~\bibnamefont{Verrocchio}},\ }%
  \bibfield{journal}{%
  \bibinfo {journal} {J. Chem. Phys.}\ }%
  \textbf{\bibinfo {volume} {136}},\ \bibinfo {pages} {204502} (\bibinfo {year}
  {2012})%
  \bibAnnoteFile{NoStop}{self:jcp12}%
\bibitem{mosaic:cammarota11}%
  \BibitemOpen
  \bibfield{author}{%
  \bibinfo {author} {\bibfnamefont{C.}~\bibnamefont{{Cammarota}}}, \bibinfo
  {author} {\bibfnamefont{G.}~\bibnamefont{{Biroli}}}, \bibinfo {author}
  {\bibfnamefont{M.}~\bibnamefont{{Tarzia}}},\ and\ \bibinfo {author}
  {\bibfnamefont{G.}~\bibnamefont{{Tarjus}}},\ }%
  \bibfield{journal}{%
  \bibinfo {journal} {Phys. Rev. Lett.}\ }%
  \textbf{\bibinfo {volume} {106}} (\bibinfo {month} {Mar.}\ \bibinfo {year}
  {2011})%
  \bibAnnoteFile{NoStop}{mosaic:cammarota11}%
\end{thebibliography}%

\end{document}